\documentclass[10pt]{article}

\usepackage{graphicx}  % needed for figures
\usepackage{amsmath}
\usepackage{multicol}
\usepackage{wrapfig}
\usepackage[left=2cm, right=2cm, top=1cm]{geometry}
\usepackage[font=it]{caption}
\usepackage{float}

\usepackage{bmpsize}

\usepackage{etoolbox}
\patchcmd\thebibliography
 {\labelsep}
 {\labelsep\itemsep=0pt\parsep=0pt\relax}
 {}
 {\typeout{Couldn't patch the command}}
\begin{document}

\title{\bf \normalsize Zonal profile corrugations and staircase formation: role of the transport crossphase}
\author{\normalsize M. Leconte$^{1,a}$ and T. Kobayashi$^{2,3}$ \\
\normalsize $^1$ National Fusion Research Institute, Daejeon 34133, South Korea \\
\normalsize $^2$ National Institute for Fusion Science, National Institutes of Natural Sciences, Toki 509-5292, Japan \\
\normalsize $^3$ The Graduate University for Advanced Studies, SOKENDAI, Toki 509-5292, Japan \\
\normalsize $^a$ corresponding author: mleconte@nfri.re.kr \\}
\date{}

\newcommand{\dif}{\partial}
\newcommand{\wk}{\omega_k}
\newcommand{\kx}{k_x}
\newcommand{\ky}{k_y}
\newcommand{\gl}{\gamma_L}
\newcommand{\dia}{\omega_{*0}}
\newcommand{\cp}{\theta_k}

\newcommand{\zv}{U}
\newcommand{\zn}{N}

%\pacs{52.35.Ra, 52.25.Fi, 52.35.Mw, 52.35.Py}

\maketitle

\begin{abstract}
Recently, quasi-stationary structures called $E \times B$ staircases were observed in gyrokinetic simulations, in all transport channels [Dif-Pradalier et al. Phys. Rev. Lett. 114, 085004 (2015)]. We present a novel analytical theory - supported by collisional drift-wave fluid simulations - for the generation of density profile corrugations (staircase), independent of the action of zonal flows: Turbulent fluctuations self-organize to generate quasi-stationary radial modulations $\Delta \cp(r,t)$ of the transport crossphase $\cp$ between density and electric potential fluctuations. The radial modulations of the associated particle flux drive zonal corrugations of the density profile, via a modulational instability. In turn, zonal density corrugations regulate the turbulence via nonlinear damping of the fluctuations.
\end{abstract}

\begin{multicols}{2}

%\section{Introduction}
High-confinement regimes provide promising conditions for future fusion devices like ITER. The H-mode, in particular has been the focus of research efforts for more than 30 years. See e.g. Refs \cite{ConnorWilson2000,Burrell2020, Kobayashi2020} for a review. The presence of turbulence-driven flows, i.e. zonal flows (ZF) have been shown to facilitate access to H-mode, by shearing apart turbulence eddies \cite{DiamondIIH2005}. Patterns are ubiquitous in non-equilibrium systems \cite{CrossGreenside2009}.
The most common radial pattern first observed in gyrokinetic simulations of ion-temperature gradient driven (ITG) turbulence has been dubbed `$E \times B$ staircase' \cite{DifPradalierHornungGhendrih2015, DifPradalierHornungGarbet2017, KosugaDiamond2014}, due to its quasi-periodic nature, for which the leading explanation is that zonal flows are responsible for the pattern, and that the density and temperature profile corrugations and hence transport modulation are a consequence of the zonal flow pattern directly suppressing the turbulence intensity \cite{AshourvanDiamond2017,GuoDiamondHughes2019}.
Similar patterns were observed in the KSTAR tokamak and reproduced by global $\delta f$ gyrokinetic simulations of collisionless trapped-electron modes (CTEM) \cite{Choi2019,LeiKwonHahm2019}.
However, it is well-known that turbulent transport does not only depend on the turbulence intensity but also on the phase-angle, i.e. transport crossphase \cite{TynanFujisawaMcKee2009, Kobayashi2017, Camargo1995} between electric potential and the advected quantity driving the turbulence. \\
\indent The turbulent particle flux can be written in the form:
$\Gamma = \displaystyle \sum_k k_y \sqrt{|n_k|^2} \sqrt{|\phi_k|^2} \gamma_{\rm coh}^2 \sin \cp$. Here, $|n_k|^2$ and $|\phi_k|^2$ denote the power spectrum of density and potential fluctuations, respectively, $\gamma_{\rm coh}$ is the square coherence, assumed here to be $\gamma_{\rm coh} \simeq 1$ for simplicity, and $\cp$ is the transport crossphase (crossphase spectrum), i.e. the phase-angle between density and potential fluctuations.
In the litterature of analytical models and 1.5D transport models, it is widely assumed that the transport crossphase between density and potential is linear, leading to the so-called `$i \delta$' prescription. There are some exceptions, e.g. Refs \cite{Terry2003, WareTerryDiamond1996}. In gyrokinetic simulations, the nonlinear crossphase in wavenumber space appears to closely match its linear value, supporting the $i \delta$ prescription for ITG and TEM turbulence \cite{ToldJenkoGorler2013}. Gyrokinetic simulations of collisionless trapped-electron mode \cite{LangParkerChen2008} revealed that - in certain parameter regimes e.g. cold ions relative to electrons - zonal flows are ineffective at suppressing turbulence, and instead zonal density generation becomes the dominant saturation mechanism. In Ref. \cite{LeconteSingh2019}, one of the authors (M.L) investigated the dynamics of the transport crossphase in dissipative trapped-electron mode turbulence, using a parametric instability approach. \\
\indent In this work, we show some evidence that the transport crossphase might in fact be nonlinear, but that this nonlinearity is manifest mostly via radial modulations of the crossphase, not predicted by linear theory. This radial modulation is responsible for the generation of zonal density corrugations. Such corrugations of the density profile were observed in Ref. \cite{Kobayashi2014} during limit cycle oscillations preceding the L-H transition in JFT-2M, using the heavy ion beam probe diagnostic (HIBP), although another possible explanation is turbulence spreading.
Here are the main findings of this work:
i) The present theory  takes into account the convective $E \times B$ nonlinearity, and thus goes beyond the well-known `$i \delta$' quasi-linear approximation, ii) This nonlinear mechanism conserves energy between turbulence and zonal density. Since zonal density is a radial mode - with poloidal and toroidal wavenumbers $m=0, n=0$ - it cannot drive transport and thus provides a benign reservoir of energy for the turbulence, and iii) In fluid simulations of collisional drift-wave turbulence, the radial modulation of the transport crossphase and associated staircase pattern of zonal density have been confirmed, and are shown to partially stabilize turbulence.

%\section{Model}
We analyse the 2 field modified Hasegawa-Wakatani model, a basic representative model for edge turbulence in magnetized plasmas \cite{WakataniHasegawa1984, Numata2007,StoltzfusDueck2016}:
\begin{eqnarray}
\frac{\dif n}{\dif t} + \{ \phi, n\} + \kappa \frac{\dif \phi}{\dif y} & = & - \alpha ( \tilde n - \tilde \phi),
\label{hw1} \\
\frac{\dif \nabla_\perp^2 \phi }{\dif t}  + \{ \phi, \nabla_\perp^2 \phi \} & = & - \alpha (\tilde n - \tilde \phi),
\label{hw2}
\end{eqnarray}
where $n$ is the electron density, $\phi$ is the electric potential, $\{ f , g \} = \frac{\dif f}{\dif x} \frac{\dif g}{\dif y}  - \frac{\dif f}{\dif y} \frac{\dif g}{\dif x}$ denote Poisson brackets and $x,y$ denote local radial and poloidal directions in  a tokamak, respectively. Here, $\tilde n = n - \langle n \rangle$ and $\tilde \phi = \phi - \langle \phi \rangle$ denote the non-zonal components of the fields, and $\langle \ldots \rangle = (1/L_y) \int \ldots dy$ is the zonal average. The quantity $\kappa = \rho_s/ L_n$ is the normalized density-gradient with $L_n = n_0 /|\nabla n_0|$, $\rho_s = c_s / \omega_{c,i}$ the sound gyroradius, $c_s = \sqrt{T_e / m_i}$ the sound speed, and $\omega_{c,i} = eB / m_i$. The coefficient $\alpha = k_\parallel^2 v_{Te}^2 / \nu_{ei}$ is the adiabaticity parameter, with $k_\parallel \sim 1 / (qR)$ the parallel wavenumber, $q$ the safety factor, $R$ the major radius, $v_{Te} = \sqrt{T_e / m_e}$ the electron thermal velocity and $\nu_{ei}$ the electron-ion collision frequency. Time and space are normalized as: $\omega_{c,i}t \to t$ and $\rho_s \nabla_\perp \to \nabla_\perp$.

\begin{figure}[H]
\begin{center}
\includegraphics[width=0.8\linewidth]{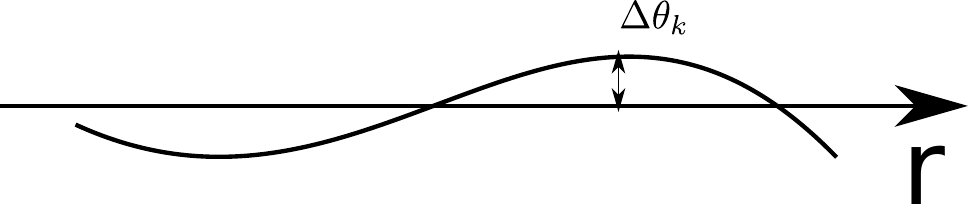}
\caption{Cartoon showing the radial modulation $\Delta \cp$ of the transport crossphase,  induced by zonal profile corrugations.}
\label{radmod}
\end{center}
\end{figure}

We extend the wave-kinetic formalism \cite{MattorDiamond1994,Bretherton1969,Sasaki2018}, to include zonal profile corrugations, i.e. zonal density.
%We extend the wave-kinetic model of Sakaki \emph{et al.} \cite{Sasaki2018}, to include zonal profile corrugations, i.e. zonal density.
For collisional drift-waves $\alpha > \wk$, linearizing Eq. (\ref{hw1}) yields:
$
n_k^L = \left[ 1 - i  \frac{\dia - \wk^L}{\alpha} \right] \phi_k
$
which provides the linear density response:
$
n_k^L = ( 1 - i \cp^0 ) \phi_k,
$
with $\cp^0= (\dia - \wk^L)/ [(1+ k_\perp^2)\alpha]$ the linear transport crossphase.
Moreover, the first-order modulation of Eq. (\ref{hw1}) yields:
$
\Delta n_k = i (\dia - \wk^L) \frac{ \ky \nabla_x \zn }{ (1+ k_\perp^2)\alpha\dia} \phi_k,
$
which provides the nonlinear correction to the density response:
$
\Delta n_k = - i \Delta \cp \phi_k,
$
with $\Delta \cp \simeq - \ky \nabla_x \zn/ \alpha$ the crossphase radial modulation. Here $\nabla_x \zn(x,t)$ denotes the zonal density gradient. Note that the zonal flows are neglected in the nonlinear density response. The coupling of zonal density to zonal flows could be important, but this is left for future work.
Hence, in the weak-turbulence approximation \cite{ZhouZhuDodin2019}, the nonlinear density response is:
$n_k \simeq  [1- i \cp^0 -i \Delta \cp(x,t) ] \phi_k$, with $\Delta \cp / \cp^0 = - \nabla_x N / |\nabla_x n_0|$, where $\nabla_x n_0 < 0$ is the equilibrium density gradient.
Physically, this means that the transport crossphase is \emph{nonlinear} - due to the $E \times B$ convective nonlinearity - and can be viewed as a radial modulation of the crossphase due to zonal profile corrugations. This physical mechanism is sketched [Fig \ref{radmod}].

Substracting Eq. (\ref{hw2}) from Eq. (\ref{hw1}) yields the conservation of potential vorticity i.e. gyrocenter ion density:
$
\frac{\dif}{\dif t} (n -\nabla_\perp^2 \phi) + v_{*0} \frac{\dif \phi}{\dif y} + {\bf v}_E . \nabla (n - \nabla_\perp^2 \phi) = 0.
$
Using the weak-turbulence approximation \cite{ZhouZhuDodin2019}, this can be written:
$
\frac{\dif}{\dif t} (n_k + k_\perp^2 \phi_k) + i \dia \phi_k + {\bf V}_{zon} . \nabla (n_k + k_\perp^2 \phi_k) = 0,
$
where ${\bf V}_{zon} = \hat z \times \nabla \phi_{zon}$ denotes zonal flows.
After some algebra, this can be written in the form of a Schrodinger-like equation:
$
i \frac{\dif}{\dif t} \Big[(R_k + k_\perp^2) \phi_k \Big] = \Big[ \dia + \ky U (R_k + k_\perp^2) \Big] \phi_k ,
$
with $R_k = 1- i (\cp^0 + \Delta \cp)$, and $U = V_{zon}$. \\
Assuming $|\dif_t \Delta \cp| \ll |\dif_t \phi_k|$, valid for slow crossphase modulation, one obtains:
$
i \frac{\dif \phi_k}{\dif t}   = \Big[ \frac{\dia}{1 + k_\perp^2 - i (\cp^0 + \Delta \cp) } + \ky U(x,t)  \Big] \phi_k.
$
Using the approximation $|\cp^0|, |\Delta \cp| \ll 1$, this reduces to:
$
i \frac{\dif \phi_k}{\dif t}   = H_H \phi_k +i H_A \phi_k,
$
where $H_H = \wk + \ky U(x,t)$ and $H_A= \gamma_k^0 + \Delta \gamma_k(x,t)$ denote the Hermitian and anti-Hermitian parts of the `Hamiltonian', respectively.
Here $\gamma_k^0 = \wk^L \cp^0$ is the linear growth-rate associated to the linear crossphase $\cp^0$, and $\Delta \gamma_k = \wk^L \Delta \cp(x,t)$ is the nonlinear growth-rate associated to the nonlinear part $\Delta \cp(x,t) \sim - \ky \nabla_x N$ of the crossphase.
Following Ref.\cite{ZhouZhuDodin2019}, one obtains the following wave-kinetic equation:
$
\frac{\dif W_k}{\dif t} + \{ H_H, W_k \}  = 2 H_A  W_k
$
where here $\{ \cdot, \cdot \}$ denotes the Poisson bracket in $(k_x, x)$ extended phase space.
After some algebra, one obtains the following reduced model:
\begin{align}
\frac{\dif W_k}{\dif t} + \frac{\dif \wk}{\dif \kx} \nabla_x W_k - \ky \nabla_x \zv \frac{\dif W_k}{\dif \kx} = 2 \gl W_k \notag\\
-  2 c_k W_k \nabla_x \zn
- \Delta\omega W_k^2,
\quad \label{wke1} \\
\frac{\dif \zv}{\dif t} = - \nabla_x \Pi + \nu_\perp \nabla_{xx} \zv - \mu \zv,
\label{zv1} \\
\frac{\dif \zn}{\dif t} = - \nabla_x \Gamma + \nabla_x \Big[ D_0 \nabla_x \zn \Big],
\qquad \label{zn1}
\end{align}
%Details of the derivation are given in Appendix. \\
Eq. (\ref{wke1}) is the extended wave kinetic equation (EWKE) for the wave action density $W_k$ , which includes a nonlinear contribution to the growth-rate, due to the zonal density induced modulation of the transport crossphase, second term on the r.h.s. of Eq. (\ref{wke1}). Eqs. (\ref{zv1}) and (\ref{zn1}) describe the dynamics of zonal flows $\zv$ and zonal density corrugations $\zn$, respectively. The frequency $\wk$ is the nonlinear frequency, including Doppler-shift due to zonal flows:
$
\wk = \frac{\dia}{1+ k_\perp^2} + \ky \zv, %- \ky c_s \nabla_x \zn,
$
with $k_\perp^2 = \kx^2 + \ky^2$,
and $W_k$ denotes the wave-action density for drift-waves with adiabatic electrons, given by:
$
W_k = (1+ k_\perp^2)^2 |\phi_k|^2,
$
with $|\phi_k|^2$ the turbulent power spectrum, and $D_0$ is a diffusivity that includes residual turbulence and neoclassical contributions.
Moreover, the nonlinear growth-rate, due to non-adiabatic electrons is:
$
\gamma_{k} = \gl -c_k \nabla_x \zn,
$
with $\gl = \cp^0 \wk^L$ the linear growth-rate.
Here, the coefficient $c_k = c_s \ky \cp^0 / (1+ k_\perp^2)$ in Eq. (\ref{wke1}) determines the coupling of the zonal density gradient to the microturbulence.

\begin{figure}[H]
\includegraphics[width=0.5\linewidth]{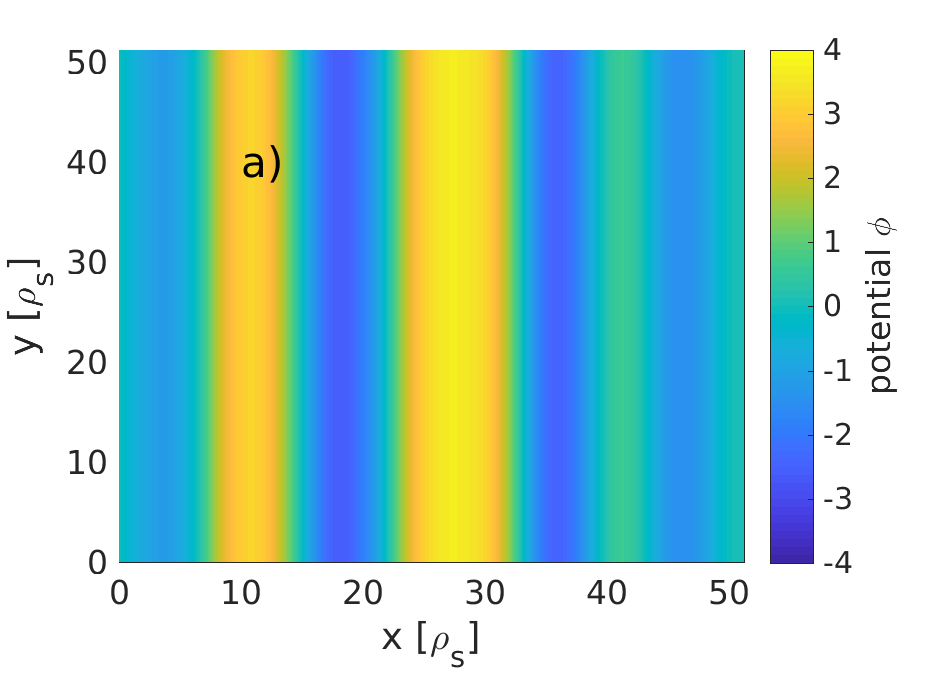}\includegraphics[width=0.5\linewidth]{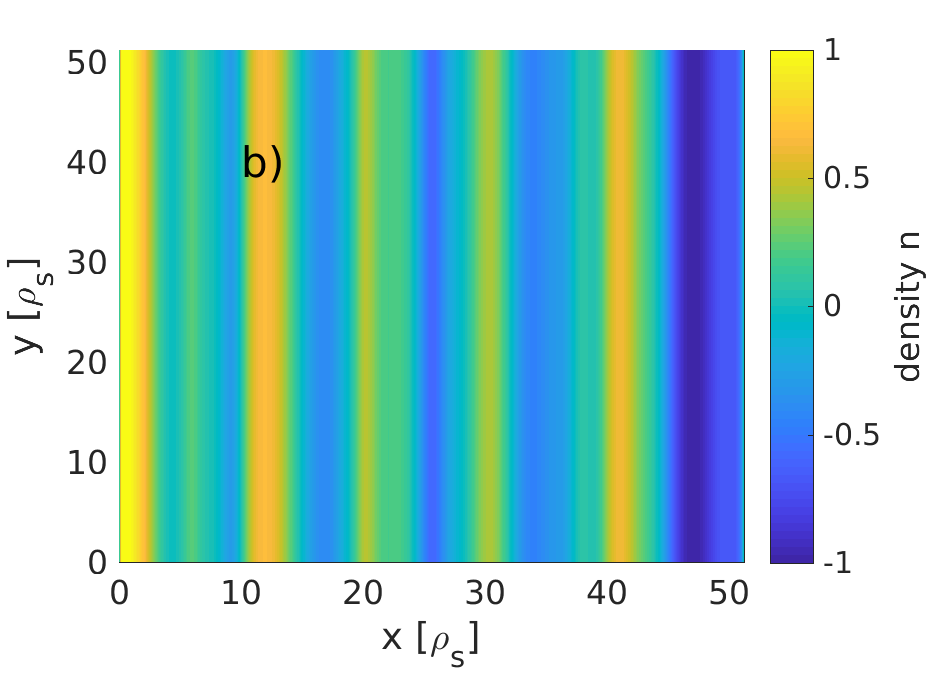}
\caption{Both zonal flows and zonal density corrugations are turbulence-driven: snapshots of a) potential and b) density in the saturated state ($t=1500$), including zonal components.}
\label{fig-snapshot}
\end{figure}

\begin{figure}[H]
\begin{center}
\includegraphics[width=0.8\linewidth]{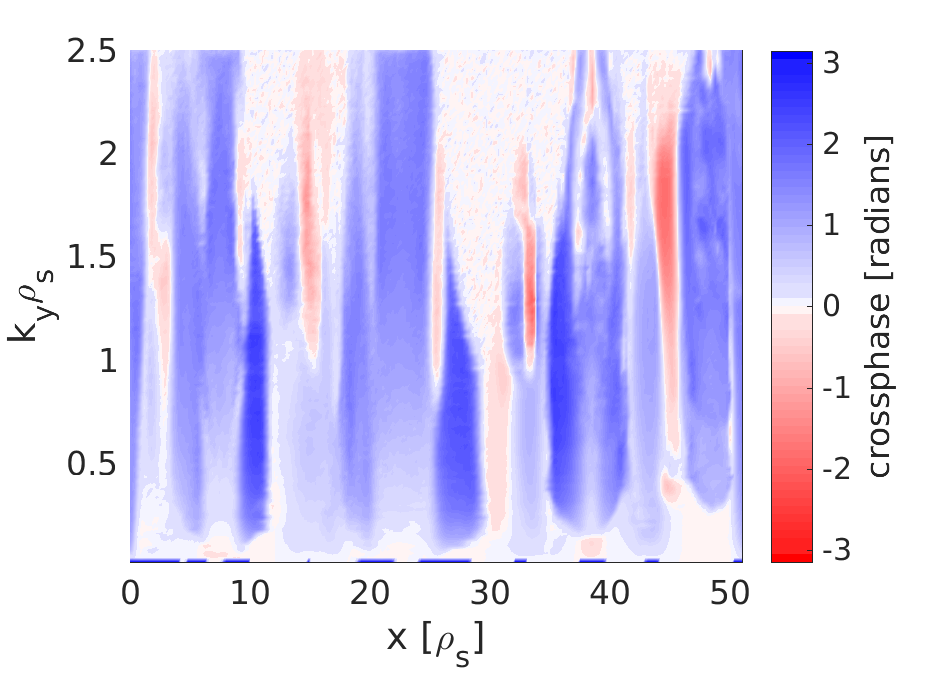}
\caption{The transport crossphase is radially-modulated: time-averaged nonlinear crossphase spectrum $\cp(x)$, v.s. wavenumber $\ky$ and radial position $x$.}
\label{fig-radmod}
\end{center}
\end{figure}

We now describe the model. The first term  on the r.h.s. of Eq. (\ref{wke1}) is the turbulence drive with linear growth-rate $\gl({\bf k})$, the second term on the r.h.s. is the nonlinear contribution to the growth-rate, proportional to the zonal density gradient $\nabla_x \zn$. %where $c_k$ is the $k$-dependent proportionality coefficient.
The first term on the r.h.s. of Eq. (\ref{zv1}) is the Reynolds torque which involves the Reynolds stress $\Pi = \langle \tilde v_x \tilde v_y \rangle$. The first term on the r.h.s. of Eq. (\ref{zn1}) is the convective particle flux $\Gamma = \langle \tilde v_x \tilde n \rangle = \sum_k (i c_s \ky/2) [n_k \phi_k^* -n_k^* \phi_k]$.
The Reynolds stress can be expressed in the form:
$
\Pi = - \sum_{\ky} \int \frac{\ky \kx W_k }{(1+ k_\perp^2)^2} d \kx.
$
Moreover, the particle flux can be approximated as:
$
\Gamma \simeq \Gamma_0 + \Delta \Gamma,
$
where $\Gamma_0= c_s \sum_{\ky} \int \frac{  \ky \cp^0 W_k }{(1+k_\perp^2)^2} d \kx$ is the particle flux due to the linear crossphase, and $\Delta\Gamma= c_s \sum_{\ky} \int \frac{  \ky \Delta\cp W_k }{(1+k_\perp^2)^2} d \kx$ is the contribution due to the radial crossphase modulation.
Here, $\cp^0 = \gl / \wk^L$ is the linear crossphase, $\wk^L = \omega_*^0 / (1+ k_\perp^2)$ is the linear DW frequency, $\Delta \cp = - \cp^0 \ky \nabla_x \zn / \dia$ is the radial crossphase modulation, and the approximation $\sin (\cp^0+ \Delta \cp) \simeq \cp^0 + \Delta\cp$ was used, since we assume $|\cp^0 + \Delta\cp| \ll 1$. \\
% ENERGY TRANSFER
Multiplying Eq.(\ref{wke1}) by $\wk^L$ and applying $\sum_{\ky} \int d \kx$, one obtains the evolution of the turbulence energy density $E_{turb} = \sum_{\ky} \int d \kx (1+k_\perp^2) |\phi_k|^2$. Multiplying Eq.(\ref{zn1}) by $\zn$, one obtains the evolution of the energy density of zonal density $E_\zn=\zn^2$.
Note that the energy is conserved in the turbulence - zonal density interaction, up to  dissipative effects due to $\Delta \Gamma$, since $\int (T_{turb}^N + T_N) dx = 0$, where $T_{turb}^N = -2 \Gamma_0 \nabla_x \zn$, and $T_N = -2 N \nabla_x \Gamma_0$ represent energy transfer terms associated to $E_{turb}$ and $E_N$, respectively. We stress out that this arises independently from the well-known energy conservation in the turbulence - zonal flow interaction. This is remarkable as it opens the way to \emph{transport decoupling} in more sophisticated models, e.g. the possibility of different magnitude of profile corrugations for different channels such as particle transport channel and thermal transport channel. \\
% FLUID SIMULATIONS
%\section{Evidence of the proposed mechanism in numerical simulations of HW turbulence}
%\subsection{Numerical results} 
To test the main hypothesis underlying the present theory, i.e. whether or not radial modulations of the transport crossphase are generated by turbulence, fluid simulations of collisional drift-wave turbulence described by the Hasegawa-Wakatani model  (\ref{hw1},\ref{hw2}) were performed using the BOUT++ framework \cite{Dudson2009}, employing the solver `'PVODE'' with adaptative time stepping to advance in time.
The model used \cite{model-used} is 2D with a resolution of $256^2$, integrated over a square of length $L=51.2$. The adiabaticity parameter is set to $\alpha=1$, and the equilibrium density gradient is $\kappa=0.5$.
Simulations are carried out until the turbulence saturates and a statistically stationary state is reached. Snapshots of potential contour [Fig.\ref{fig-snapshot}a] and density contour [Fig.\ref{fig-snapshot}b] are shown, including zonal components. Contours of potential and density are elongated in the poloidal direction $y$ due to zonal flows and zonal density, respectively.
In the saturated state, the nonlinear transport crossphase $\cp = \arg ( n_k^* \phi_k)$, averaged over time, is shown v.s. poloidal wavenumber $\ky$ and radial direction $x$ [Fig.\ref{fig-radmod}]. Here, $\arg(z)$ denotes the argument of the complex $z$. A modulation pattern is clearly observed in the radial direction $x$ [Fig.\ref{fig-radmod}]. To our knowledge, this is the first time that such a radial modulation of transport crossphase has ever been observed in numerical simulations. However, more work needs to be done to investigate the effects of the simulation parameters $\alpha$ and $\kappa$ on the transport crossphase modulation.

\begin{figure}[H]
\begin{center}
\includegraphics[width=0.8\linewidth]{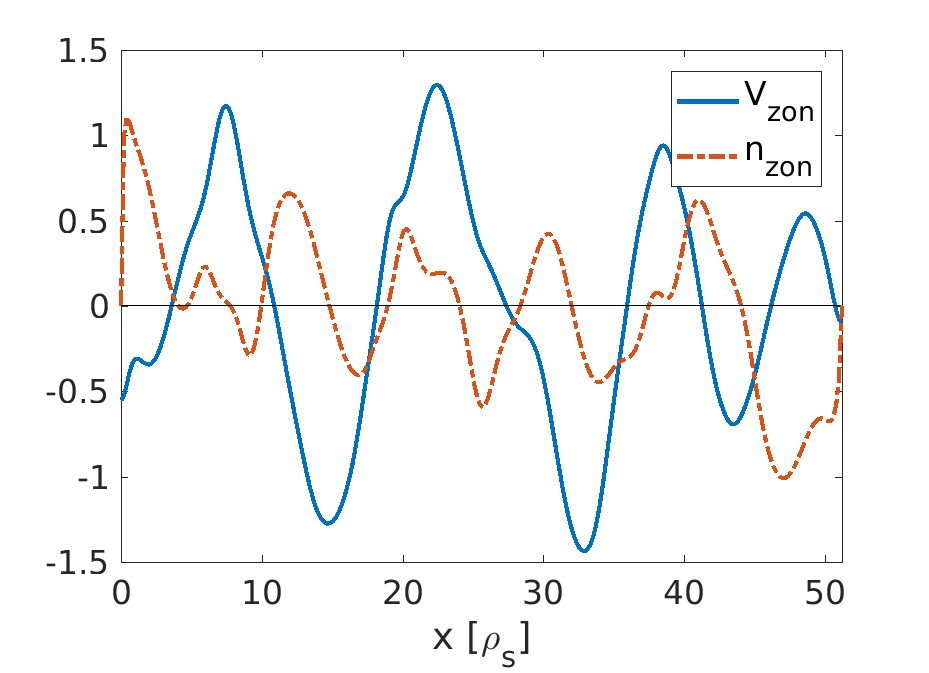}
\caption{Zonal flows and zonal density have different scale: time-averaged profiles of zonal flows $V_{\rm zon}$ (solid, blue) and zonal density $n_{\rm zon}$ (dash, red).}
\label{fig-prof}
\end{center}
\end{figure}

\begin{figure}[H]
\begin{center}
\includegraphics[width=0.8\linewidth]{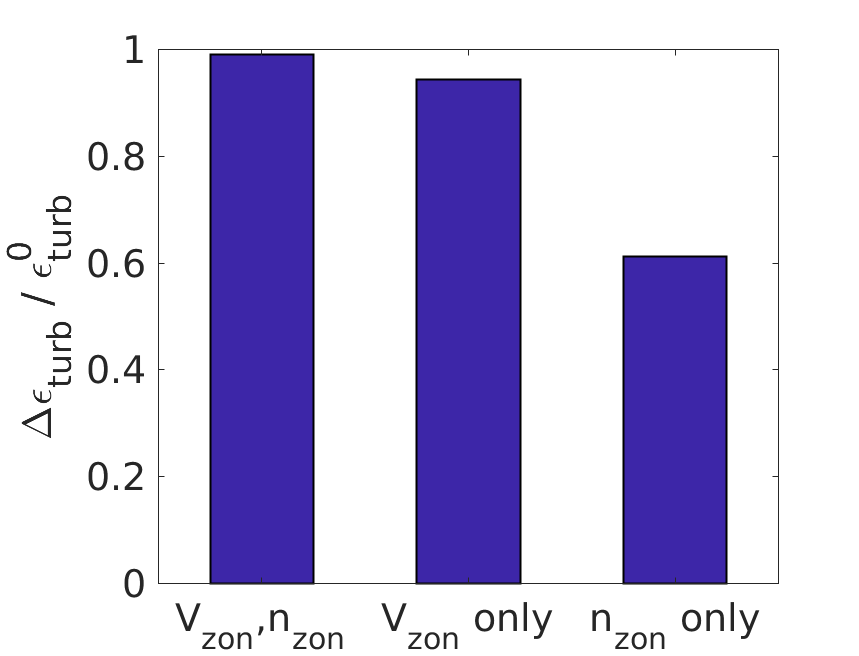}
\caption{Zonal flows and zonal density can both suppress turbulence: zonal efficiency $\Upsilon = \Delta \epsilon_{turb} / \epsilon_{turb}^0$ for different cases, 1st bar: with both zonal flows and zonal density, 2nd bar: with artificially-suppressed zonal density, 3rd bar: with artificially-suppressed zonal flows.}
\label{fig-effzon}
\end{center}
\end{figure}

The profiles of zonal flows and zonal density, averaged over time,  are shown for simulations parameters $\alpha=1$, $\kappa=0.5$ [Fig.\ref{fig-prof}]. It is apparent from Fig.\ref{fig-prof} that the zonal density profile evolves on a much-smaller scale than that of zonal flows. This is somewhat surprising and not intuitive, since in the litterature, zonal flows and zonal density are described as two components of `zonal modes' with the same wavenumber ${\bf q}=q_x \hat x$. Hence, if zonal modes were really describable in this way, we would naively expect that they would have approximately the same scale, i.e the same dominant wavenumber. The fact that they have widely different scale - and hence different dominant wave number - points to the intrinsically nonlinear nature of these fields. This seems to be a generic feature in the Hasegawa-Wakatani model (in the adiabatic electron regime $\alpha>\kappa$ where zonal modes are important), as it is observed also in simulations with $\alpha=10, \kappa=0.5$ and with $\alpha=1, \kappa=0.2$ (not shown). The different scale between zonal flows and zonal density in drift-wave turbulence was first observed in Ref. \cite{KimAn2019}. This difference of scale between zonal flows and zonal density can be interpreted as a \emph{decoupling} between different transport channels, i.e. vorticity transport v.s. particle transport. For a more sophisticated model, this may have implications for the important phenomenon of \emph{transport decoupling}, e.g. between particle transport v.s. heat transport, but more work needs to be done to confirm this picture. Let us point out one possible consequence of the scale difference of zonal density v.s. zonal flows. This could mean a suppression mechanism due to zonal flows for larger scale eddies v.s. one due to zonal density for small scale eddies.  \\
% COMPARISON
%\subsection{Comparison with the extended wave-kinetic model \& with JFT-2M experimental data}
%\subsubsection{Comparison with the extended wave-kinetic model}
The extended wave-kinetic model predicts that zonal flows and zonal density corrugations both play a role in  transport suppression. Hence, one should not be dominant over the other. This is consistent with the time-average profile of Fig.\ref{fig-prof} which shows them to be roughly of the same magnitude. The reduced model also predicts the radial modulation $\Delta \cp(x,t)$ of the transport crossphase, which is confirmed in Fig.\ref{fig-radmod}.
We verified that crossphase modulations vanish completely when both zonal flows and zonal density are artificially suppressed (not shown). This implies that this radial modulation is a nonlinear phenomenon and is not predicted by quasi-linear theory which uses the linear `$i \delta$' prescription for the electron response.
One may ask: What is the qualitative effect of zonal density v.s. zonal flows in suppressing the turbulence?
To investigate this, we compare the turbulence level, i.e. the time-averaged turbulence energy at saturation, for different cases. It is convenient to introduce a normalized indicator: the \emph{zonal efficiency} $\Upsilon$ (\%), defined as:
$
\Upsilon = \frac{\Delta \epsilon_{turb}}{\epsilon_{turb}^0},
$
where $\Delta \epsilon_{turb} = | \epsilon_{turb} - \epsilon_{turb}^0 |$, $\epsilon_{turb}$ denotes the time-averaged turbulence energy at saturation, and $\epsilon_{turb}^0$ is its reference value when both zonal flows and zonal density are artificially-suppressed.
The zonal efficiency indicates how strongly different zonal structures are able to suppress turbulence. It is shown for different cases [Fig.\ref{fig-effzon}]. One observes that the case with zonal flows and zonal density has a zonal efficiency of $\Upsilon \sim 99.1\%$ close to 100\%, corresponding to almost totally suppressed turbulence. The case with artificially-suppressed zonal density has a zonal efficiency of $ \Upsilon \sim 94.4\%$, thus lower than the case with both zonal flows and zonal density present, although not by a large margin. However, the most interesting case is the one with zonal density alone, i.e. with artificially-suppressed zonal flows, with a zonal efficiency of $\Upsilon \sim 61.2\%$,which is still large. This shows that zonal density corrugations may play an important role in turbulence suppression in some regimes. Note that the sum of the two rightmost columns in Fig. \ref{fig-effzon} do not add up to $100\%$. That is each turbulence-suppression mechanism (zonal flows v.s. zonal density) separately suppresses most of the turbulence. This could mean that either the scales of the two mechanisms - being different - overlap in most of the intermediate scales, or that the interaction of the two mechanisms is non-linear.

%\subsubsection{Qualitative comparison with JFT-2M experimental data}
We also identify similar qualitative features of the theory with experimental data from a previous JFT-2M experiment investigating limit-cycle oscillations (LCO) during the L-H transition \cite{Kobayashi2014}. Data from the heavy ion beam probe (HIBP) shows qualitative features - namely zonal density corrugations - consistent with the main characteristics of our model, although turbulence spreading is another possible explanation as discussed in Ref. \cite{Kobayashi2014}. We leave detailed comparison for future work.
Fig. 11 in Ref. \cite{Kobayashi2014} shows a slow radial modulation of the HIBP signal intensity profile, a proxy for electron density.
This edge density perturbation excites inward propagating zonal density pulses, which are poloidally symmetric, as discussed in Fig. 18 (a) in Ref. \cite{Kobayashi2014}.
The sound Larmor radius is $\rho_s \sim \rho_i \sim 1.2 mm$ in this experiment, where $T_i \sim T_e$ is assumed and $\rho_i$ is the ion Larmor radius. The assumption $T_i \sim T_e$ is verified in JFT-2M \cite{IdaHidekumaMiura1990}. Our model assumes cold ions $T_i \ll T_e$, hence we can only make qualitative comparisons. Ref. \cite{Kobayashi2014} estimated the radial wavenumber of the zonal density perturbation as $q_r \sim 25 m^{-1}$ and that of the microturbulence as $k_r \sim 10^2 m^{-1}$. This gives $q_r \rho_s \sim 0.03$ and $k_r \rho_s \sim 0.12$, hence $q_r \rho_s \ll k_r \rho_s$, consistent with a slow radial modulation.

%\begin{figure}[H]
%\begin{center}
%\includegraphics[width=0.5\linewidth]{hibp-a}\includegraphics[width=0.5\linewidth]{hibp-zon-b}
%\caption{Zonal density corrugations are observed experimentally: Experimental data from JFT-2M showing zonal density corrugations: a) original data, and b) zonal density corrugations, i.e. data with substracted mean component.}
%\label{fig-hibp}
%\end{center}
%\end{figure}

%\section{Discussion}
First, let us discuss our results concerning the extended wave-kinetic model Eqs.(\ref{wke1},\ref{zv1},\ref{zn1}). This model is an extension of the well-known wave-kinetic equation \cite{MattorDiamond1994, Bretherton1969}, to self-consistently include the physics of the transport crossphase. We can compare this model with that of Ref. \cite{Sasaki2018}. The main difference is the presence of the nonlinear part of the growth-rate in our model, second term on the r.h.s. of Eq. (\ref{wke1}), which can be traced to the convective $E \times B$ nonlinearity. This couples to the dynamics of zonal density corrugations, providing a new feedback loop which is absent in the standard wave-kinetic equation. In Ref. \cite{Sasaki2018}, the wave-kinetic model is solved numerically in the extended phase-space ($x, k_x$). It shows a complex interplay between turbulence and zonal flows that lead to nonlinear structures (patterns), associated to the `trapping' of turbulence wave-packets in the troughs of zonal flows. This is beyond the scope of this article and left for future work. Instead, we provide evidence of the validity of the model by using direct numerical simulations in real space. Nonetheless, we expect similar nonlinear structures to arise. 
Now, we discuss the zonal density generation mechanism, Eq.(\ref{zn1}). In Ref. \cite{LangParkerChen2008}, it was shown that CTEM turbulence can saturate via nonlinear generation of zonal density. Although it is not directly relevant to the present work, since we consider collisional drift-waves, we can compare with the zonal density drive mechanism described by equation (8) in \cite{LangParkerChen2008}, as this should be model-independent. Our Eq.(\ref{zn1}) differs from the one in \cite{LangParkerChen2008}, since we show that energy is conserved between turbulence and zonal density, whereas the analysis in Ref. \cite{LangParkerChen2008} is valid only for the initial exponential growth of the modulational instability. The fluid model (\ref{hw1},\ref{hw2}) could possibly be extended to CTEM, where zonal density generation seems to play a crucial role.
Another point is about the importance of density nonlinear structures. In Refs. \cite{KosugaHasamada2018, Kin2019}, it was shown that the streamer flows and associated density component may play an important role in large scale radial particle transport. In the theory of \cite{KosugaHasamada2018}, the density component of streamers plays an important role, by enhancing the modulational growth-rate of streamers. In this sense, the present theory of zonal density formation is complementary to the one of Ref. \cite{KosugaHasamada2018} for streamer formation, which correspond to the opposite limits of the more general `convective cell'. Since streamers are purely radial structures, they can only enhance turbulent transport. Hence, when discussing streamers, it may be important to consider zonal density corrugations as well since the latter are shown to suppress turbulent transport.
Another point is that the drift-fluid model used in this work is a basic representative model for edge turbulence. This is helpful to obtain a qualitative understanding of the nonlinear physical mechanisms at play. However, in order to make quantitative comparisons with tokamaks, gyrokinetic simulations (or at least gyro-Landau-fluid simulations) would be best. Specifically, it would be interesting to analyze the transport crossphase in gyrokinetic simulations of fusion experiments with transport staircases.

There are limitations to our model. First, the model assumes cold ions, $T_i \ll T_e$, and hence does not contain finite-ion Larmor radius (FLR) effects, although it includes ion inertia ($\rho_s$) effects. It is thus not directly applicable to the important ion-temperature-gradient driven mode (ITG). Second, electron temperature gradient effects ($\eta_e$) are neglected. This is beyond the scope of this article and left for future work, where we plan to investigate the possible decoupling between particle transport and thermal transport (transport decoupling).

%\section{Conclusion}
In this work, we derived the extended wave-kinetic equation (\ref{wke1}), self-consistently coupled to the dynamics of zonal flows and \emph{more importantly} of zonal density corrugations. The latter - and generalization to other profile corrugations - may be an important missing piece in the understanding of turbulent transport in fusion devices. The theory can be summarized as follows: Turbulent fluctuations self-organize to generate quasi-stationary radial modulations $\Delta \cp(r,t)$ of the transport crossphase $\cp$ between density fluctuations and potential fluctuations. This results in turbulent particle flux modulations $\tilde \Gamma(r,t)$. The radial modulation of particle flux nonlinearly drive zonal  corrugations of the density profile via a modulational instability. In turn, zonal density corrugations regulate the turbulence via nonlinear damping of the fluctuations.
%The main findings of this work are:
%i) The present theory takes into account the convective $E \times B$ nonlinearity, and thus goes beyond the well-known `$i \delta$' quasi-linear approximation,
%ii) This nonlinear mechanism conserves energy between turbulence and zonal density. Since zonal density is a radial mode ($m=0,n=0$), with $m$ and $n$ the poloidal  and toroidal mode numbers, it cannot drive transport and thus provides a benign reservoir of energy for the turbulence, and
%iii) In fluid simulations of of collisional drift-wave turbulence, the radial modulation of the transport crossphase and associated staircase profile structure have been confirmed to partly stabilize the turbulence.

%\section*{Acknowledgments}
M.L. thanks Raghvendra Singh, Jae-Min Kwon, Lei Qi, I. Dodin, Hongxuan Zhu, T. Stoltzfulz-Dueck and M.J. Pueschel for helpful discussions. M.L was supported by R\&D Program through National Fusion Research Institute (NFRI) funded by the Ministry of Science and ICT of the Republic of Korea (No. NFRI-EN2041-6).
\section*{\small Data Availability Statement}
The data that support the findings of this study are available from the corresponding author upon reasonable request.
\quad\\
\hrule
\quad\\

\begingroup
\renewcommand{\section}[2]{}

\endgroup

\end{multicols}


\begin{thebibliography}{90}
\bibitem{ConnorWilson2000}
J. Connor and H. Wilson, \emph{Plasma Phys. Control. Fusion} 42, R1 (2000).
\bibitem{Burrell2020}
K.H. Burrell, \emph{Phys. Plasmas} 27, 060501 (2020).
\bibitem{Kobayashi2020}
T. Kobayashi, \emph{Nucl. Fusion} 60, 095001 (2020).
\bibitem{DiamondIIH2005}
P.H. Diamond, S.-I Itoh, K. Itoh and T.S. Hahm, \emph{Plasma Phys. Control. Fusion} 47, R35 (2005).
\bibitem{CrossGreenside2009}
M. Cross and H. Greenside, \emph{Pattern Formation and Dynamics in Non-Equilibrium Systems (Cambridge University Press, N.Y., 2009).}
\bibitem{DifPradalierHornungGhendrih2015}
G. Dif-Pradalier, G. Hornung, Ph. Ghendrih, Y. Sarazin, F. Clairet, L. Vermare, P.H. Diamond, J. Abiteboul, T. Cartier-Michaud, C. Ehrlacher et al., \emph{Phys. Rev. Lett.} 114, 085004 (2015).
\bibitem{DifPradalierHornungGarbet2017}
G. Dif-Pradalier, G. Hornung, X. Garbet, Ph. Ghendrih, V. Grandgirard, G. Latu and Y. Sarazin, \emph{Nucl. Fusion} 57, 066026 (2017).
\bibitem{KosugaDiamond2014}
Y. Kosuga, P.H. Diamond, G. Dif-Pradalier and O.D. Gurcan, \emph{Phys. Plasmas} 21, 055701 (2014).
\bibitem{AshourvanDiamond2017}
A. Ashourvan and P.H. Diamond, \emph{Phys. Plasmas} 24, 012305 (2017).
\bibitem{GuoDiamondHughes2019}
W. Guo, P.H. Diamond, D. Hughes, L. Wang and A. Ashourvan, \emph{Plasma Phys. Control. Fusion} 61, 105002 (2019).
\bibitem{Choi2019}
M.J. Choi, H.G. Jhang, J.M. Kwon, J. Chung, M.H. Woo, L. Qi., S.H. Ko, T.S. Hahm, H.K. Park, H.S. Kim et al. \emph{Nucl. Fusion} 59, 086027 (2019).
\bibitem{LeiKwonHahm2019}
L. Qi, J.M. Kwon, T.S. Hahm, S. Yi and M.J. Choi, \emph{Nucl. Fusion} 59, 026013 (2019).
\bibitem{TynanFujisawaMcKee2009}
G.R. Tynan, A. Fujisawa and G. McKee, \emph{Plasma Phys. Control. Fusion} 51, 113001 (2009).
\bibitem{Kobayashi2017}
T. Kobayashi, K. Itoh, T. Ido, K. Kamiya, S.I. Itoh, Y. Miura, Y. Nagashima, A. Fujisawa, S. Inagaki and K. Ida \emph{Sci. Rep.} 7, 14971 (2017).
\bibitem{Camargo1995}
S. Camargo, D. Biskamp and B. Scott, \emph{Phys. Plasmas} 02, 48 (1995).
\bibitem{Terry2003}
P.W. Terry, D.E. Newman and A.S. Ware, \emph{Phys. Plasmas} 10, 1066 (2003).
\bibitem{WareTerryDiamond1996}
A.S. Ware, P.W. Terry, P.H. Diamond and B. Carreras, \emph{Plasma Phys. Control. Fusion} 38, 1343 (1996). 
\bibitem{ToldJenkoGorler2013}
D. Told, F. Jenko, T. Gorler, F.J. Casson, E. Fable and ASDEX-U team, \emph{Phys. Plasmas} 20, 122312 (2013).
\bibitem{LangParkerChen2008}
J.Y. Lang, S.E. Parker and Y. Chen, \emph{Phys. Plasmas} 15, 055907 (2008).
\bibitem{LeconteSingh2019}
M. Leconte and R. Singh, \emph{Plasma Phys. Control. Fusion} 61, 095004 (2019).
%\bibitem{LeconteDiamond2012}
%M. Leconte and P.H. Diamond, \emph{Phys. Plasmas} 19, 055903 (2012).
\bibitem{Kobayashi2014}
T. Kobayashi, K. Itoh, T. Ido, K. Kamiya, S.I. Itoh, Y. Miura, Y. Nagashima, A. Fujisawa, S. Inagaki, K. Ida et al., \emph{Nucl. Fusion} 54, 073017 (2014).
\bibitem{IdaHidekumaMiura1990}
K. Ida, S. Hidekuma, Y. Miura, T. Fujita, M. Mori, K. Hoshino, N. Suzuki, T. Yamauchi and JFT-2M Group, \emph{Phys. Rev. Lett.} 65, 1364 (1990).
\bibitem{WakataniHasegawa1984}
M. Wakatani and A. Hasegawa, \emph{Phys. Fluids} 27, 611 (1984).
\bibitem{Numata2007}
R. Numata, R. Ball and R.L. Dewar, \emph{Phys. Plasmas} 14, 102312 (2007).
\bibitem{StoltzfusDueck2016}
T. Stoltzfus-Dueck, \emph{Phys. Plasmas} 23, 054505 (2016).
\bibitem{MattorDiamond1994}
N. Mattor and P.H. Diamond, \emph{Phys. Plasmas} 1, 4002 (1994).
\bibitem{Bretherton1969}
F.P. Bretherton and C.J.R. Garret, \emph{Proc
. R. Soc. A} 302, 529 (1969).
\bibitem{Sasaki2018}
M. Sasaki, T. Kobayashi, K. Itoh, N. Kasuya, Y. Kosuga, A. Fujisawa and S.I. Itoh, \emph{Phys. Plasmas} 25, 012316 (2018).
\bibitem{ZhouZhuDodin2019}
Y. Zhou, H. Zhu and I. Dodin, \emph{Plasma Phys. Control. Fusion} 61, 075003 (2019).
\bibitem{Dudson2009}
B. Dudson, M. Umansky and X. Xu, \emph{Comput. Phys. Comm.} 180, 1467 (2009).
\bibitem{model-used}
model `hasegawa-wakatani' in the BOUT++ example folder.
\bibitem{KimAn2019}
C.B. Kim, C.Y. An and B. Min, \emph{Plasma Phys. Control. Fusion} 61, 085024 (2019).
\bibitem{KosugaHasamada2018}
Y. Kosuga and K. Hasamada, \emph{Phys. Plasmas} 25, 100701 (2018).
\bibitem{Kin2019}
F. Kin, A. Fujisawa, K. Itoh, Y. Kosuga, M. Sasaki, S. Inagaki, Y. Nagashima, T. Yamada, N. Kasuya, K. Yamasaki et al., \emph{Phys. Plasmas} 26, 042306 (2019).

\end{thebibliography}
\end{document}